# Fast directional transport of Leidenfrost droplets on spiked surfaces

Kai-Xin Hu[a,b,1], Dong-Xu Duan[a,b], Yin-Jiang Chen[a,b], Dan Wu[a,b], Qi-Sheng Chen[c,d]

[a]Zhejiang Provincial Engineering Research Center for the Safety of Pressure Vessel and Pipeline, Ningbo University, Ningbo, Zhejiang 315211, China

[b]Key Laboratory of Impact and Safety Engineering (Ningbo University), Ministry of Education, Ningbo, Zhejiang 315211, China

[c]School of Engineering Science, University of Chinese Academy of Sciences, Beijing 100190, China

[d]Key Laboratory of Microgravity, Institute of Mechanics, Chinese Academy of Sciences, Beijing 100190, China

## Abstract

The Leidenfrost effect enables droplets to levitate above a solid surface, significantly reducing the resistance to droplet motion. In this study, a spiked surface is utilized to achieve fast directional transport of Leidenfrost droplets, with a maximum average speed reaching 8.36 m/s over a 10 cm distance—far exceeding the previously reported maximum speeds for droplet transport. When a droplet falls onto a substrate heated above the Leidenfrost temperature, it becomes trapped between spikes and levitates. The sides and bottom surface of the droplet undergo vaporization, creating a gas film between the solid wall and the droplet. However, this gas film is unstable and prone to rupture at certain points, causing the droplet to come into contact with the solid surface. Therefore, the droplets undergo violent boiling, leading to intense compression and bursting into smaller daughter droplets, which are then propelled rapidly along the substrate. Additionally, the asymmetric geometry of the spikes ensures that droplets move unidirectionally along the longitudinal direction. This study proposes a novel droplet self-propulsion mechanism, pioneering new strategies for enhancing droplet transport speed.

---

[1]Corresponding author, Email: hukaixin@nbu.edu.cn

# 1. Introduction

Droplets on solid surfaces can undergo directional motion under certain driving forces. This phenomenon of droplet migration on a solid surface plays a critical role in many technological fields, such as condensation, water harvesting, microfluidic devices, and liquid transport systems, which are reviewed by many authors[1-4].

The migration of droplets on solid surfaces involves not only the interactions at the solid/liquid/gas three-phase interface, but also multiple factors such as temperature, surface structure, droplet morphology changes and surface tension. In recent years, with the advancement of technological methods and fundamental research, droplet migration can be effectively controlled through various physical mechanisms, including the wettability gradient [5], thermocapillary effect [6], Leidenfrost effect [7-12], capillary force [13], stiffness gradient [14], chemical gradient [15], and electrostatic force [16]. In Table 1, we compare the mechanisms and transport speeds of these methods.

Table 1　　Common Methods for Wall-Adhering Droplet Migration

| Mechanism | Maximum Speed | Liquid Type/Related Surface Treatment |
| --- | --- | --- |
| Stiffness gradient | $8\times10^{-8}$m/s[14] | Water/Silicone Gel Layer (PDMS) Soft Substrate |
| Capillary force | 0.078m/s[13] | Water/Superhydrophilic Surface |
| chemical gradient | 0.15m/s[15] | Water/Chemical Wetting Gradient |
| Leidenfrost effect | 0.192m/s[8] | Water/angled self-assembled microstructures |
|  | 0.6m/s[9] | Water/High-Temperature Surface with Columnar Structures |
|  | 0.05m/s[7] | Water, Ethanol, etc. / High-Temperature Surface |
|  | 0.01m/s[10] | Ethanol / flat substrate submitted to a horizontal thermal gradient |
| Electrostatic force | 1.1m/s[16] | Water, Glycerol, Ethanol, etc. / Superhydrophobic Surface with a Charge Gradient |

As shown in the Table 1, the Leidenfrost effect exhibits superior transport speed and distance. This phenomenon occurs when a liquid drop deposited on a substrate above the so-called Leidenfrost temperature $T_L$, causing rapid vaporization that forms a stable vapor cushion between the liquid and the surface. Thus, the droplet can levitate and move atop this vapor layer without direct contact with the substrate[17]. A defining feature of this effect is a minimal friction and a low heat transfer coefficient between the droplet and the hot solid surface, making it applicable to a wide range of fields, such as combustion engines [18], drag reduction [19] and power generation[20,21].

Over the last two decades, researchers have utilized asymmetric solid surfaces to achieve self-propulsion of Leidenfrost droplets. Linke et al. [7] have reported the migration of Leidenfrost droplets by hot surfaces with ratchetlike topology. Dupeux et al. [22] have shown experimentally and theoretically that asymmetric mass distributions in a Leidenfrost solid can lead to a non-homogenous vapor layer in which the lubrication flow generates a lateral force able to propel the body. Kruse et al. [8] have proposed a new mechanism for self-propelled droplets on heated surfaces with angled self-assembled microstructures. Sobac et al. [10] have theoretically investigated the behavior of self-propelled Leidenfrost drops on a flat substrate submitted to a horizontal thermal gradient. The migration speeds achieved by the aforementioned methods are listed in Table 1.

Significant progress has been made in the directional migration of Leidenfrost droplets, yet methods for controlling and guiding these highly mobile droplets remain limited. Considerable room for improvement still exists regarding migration speed and precise control.

In this study, we achieve rapid directional transport of droplets in the Leidenfrost state using millimeter-scale spiked plates, with the fastest average migration speed reaching 8.36 m/s over a 10 cm distance. We conduct experiments on plates with four different spike lengths and find that the droplets exhibit two distinct Leidenfrost states: stationary suspension and boiling-induced migration. The effects of spike length and heating temperature on the droplet migration speed and droplet size are investigated.

## 2. Experimental Setup

The experimental setup, as illustrated in Figure 1, comprises the heating stage (constant-temperature hot plate), the spiked copper plate (comprising a spike zone and a transport zone) and the High-Speed Camera (HSC).

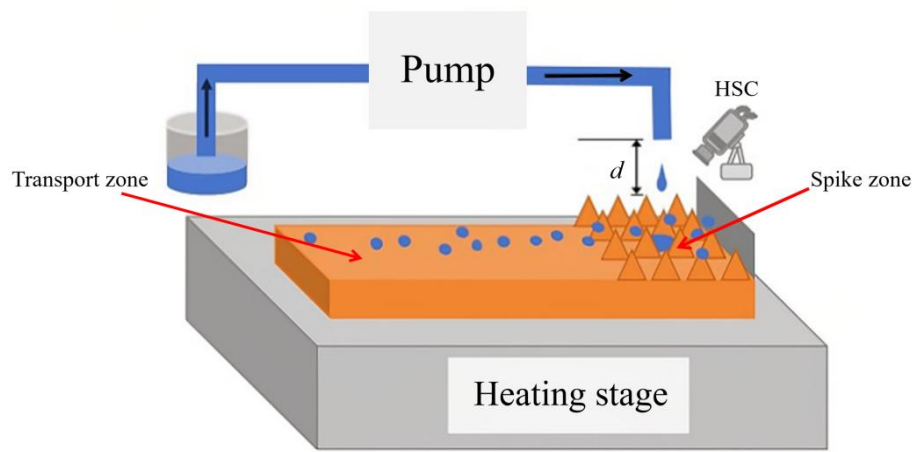

(a)

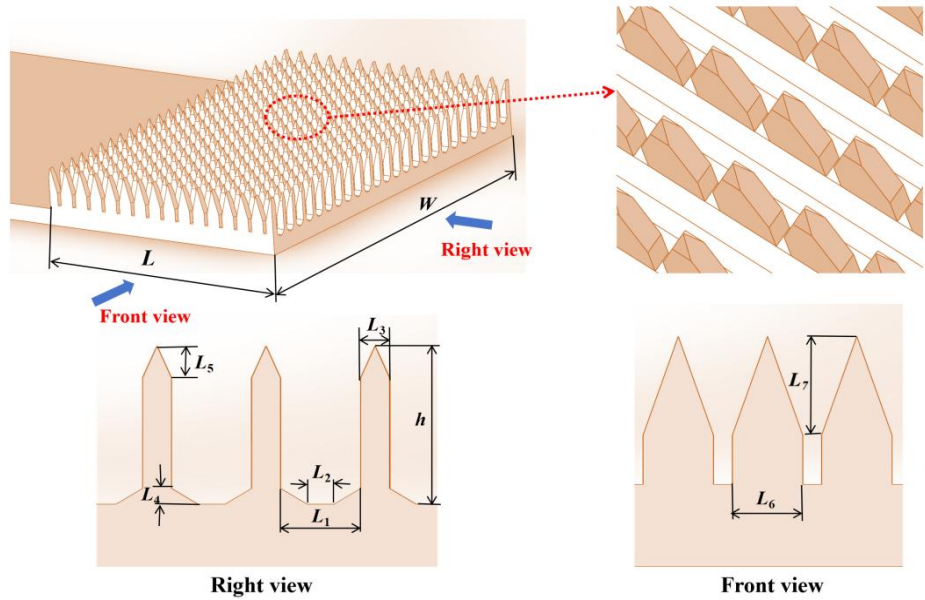

(b)

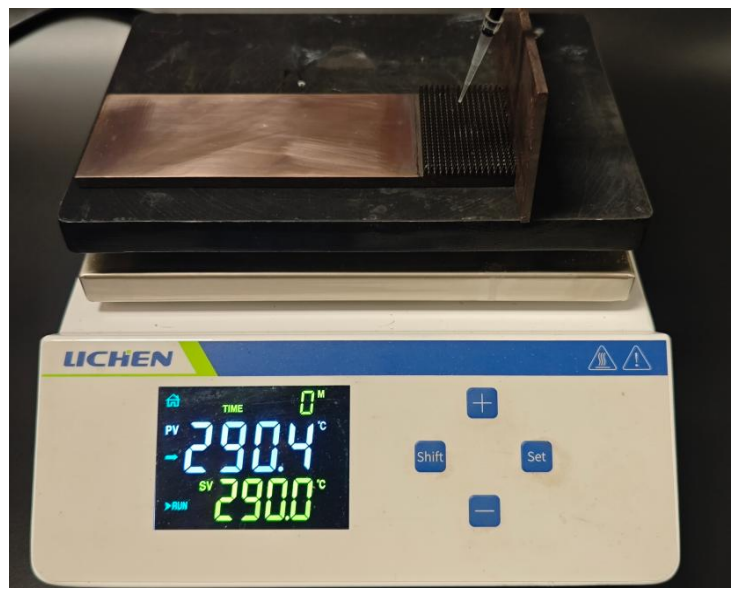

(c)

Figure 1　Experimental system: (a) schematic of experimental setup, (b) three-dimensional structure of the spiked surface and (c) physical diagram.

In the experiment, the spiked plate is placed on the heating stage within a temperature control range of 30°C~350°C. Subsequently, droplets are delivered to the spike zone using either a micropipette or peristaltic pump (with the nozzle placed at a distance of $d = 5$ mm from the spikes). The droplet boiling, bursting, and migration processes are observed by a high-speed camera. Experiments are conducted on four spiked plates with spike heights of $h$=3,5,7 and 9mm. Additional geometric parameters of the plates are provided in the appendix.

## 3. Results

3.1 Three distinct Leidenfrost states

On the flat region of the spiked plate, we measure the Leidenfrost temperature $T_L$ =(144±0.2)°C using a thermocouple. For the spiked region, we observe that droplets could levitate when the spiked plate is heated above the Leidenfrost threshold $T_L$, demonstrating the Leidenfrost phenomenon. However, due to the structural differences between the spiked plate and the flat one, the Leidenfrost droplets exhibit three distinct states.

The first one involves droplet encapsulation of the spikes, with the droplet levitating atop the spikes, as illustrated in Figure 2(a) and video 1. This case is only observed on the plate with $h$=3mm. For cases with longer spikes, the droplet tends to slide from the spikes into the microgrooves, making it unable to form a stable state where it fully envelops the spikes.

The second one occurs when droplets settle between the spikes and levitate above the substrate surface (Figure 2(b) and video 2). This behavior can be observed across all four tested spike heights ($h$=3,5,7,9 mm).

The third scenario appears when the droplet volume exceeds a critical threshold $V_m$, the droplet undergoes violent boiling, bursting into multiple daughter droplets that migrate away (Figure 2(c) and video 3). This process continues until the remaining droplet volume falls below the critical value. Subsequently, the droplet stabilizes within the microgrooves of the spikes, maintaining levitation before gradually evaporating.

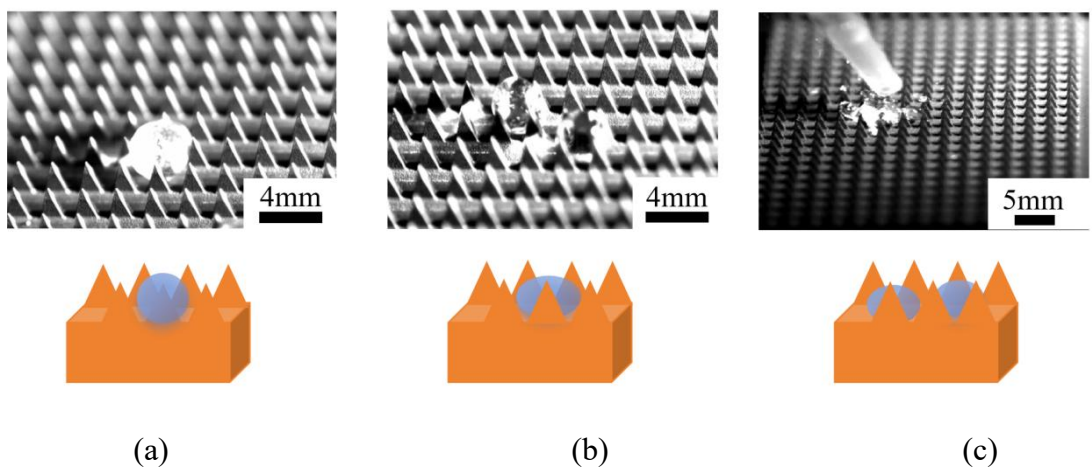

(a)　　　　　　　　　　　(b)　　　　　　　　　　　(c)

Figure 2　Three types of Leidenfrost states: (a) droplet encapsulating spikes； (b) droplet levitating between spikes; (c) droplet boiling and bursting. Here, .$h$=3mm.

The mechanism of droplet directional migration is illustrated as follows (see Figure 3). When a droplet levitates above a flat substrate, the gas film between the liquid and solid wall remains relatively stable. However, for droplets that settle between the spikes, vaporization occurs on both the side and bottom surfaces of the liquid. The evaporation at the gas-liquid interface induces a vapor recoil force on the droplet. This makes the gas film difficult to maintain stably. In particular, the gaps between adjacent spikes cause non-uniform vaporization on the droplet's side surfaces. The resultant force of vapor recoil from all interfaces makes the droplet prone to movement and oscillation, while the gas film becomes highly unstable.

When the droplet volume is relatively small (Figure 3(a) and video 4), the droplet continues to oscillate in the longitudinal direction (Figure 3(c)) while its volume gradually decreases until it vanishes. When the droplet volume exceeds $V_m$ (Figure 3(b) and video 5), such oscillation readily causes the droplet to come into contact with the solid wall (Figure 3(d)), subsequently triggering boiling and bursting.

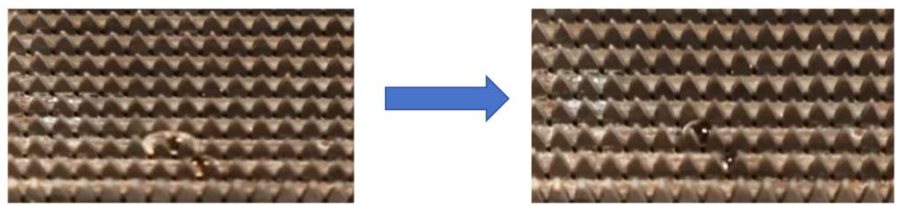

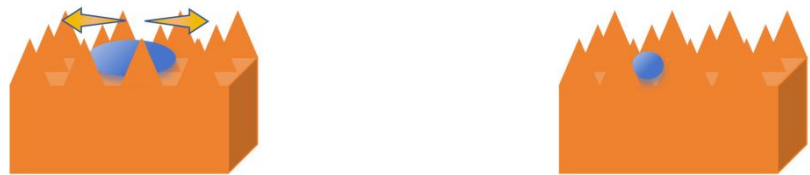

(a)

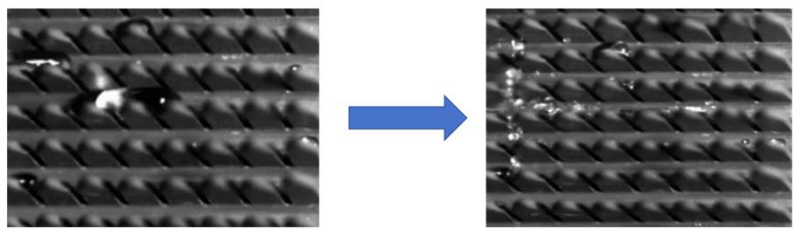

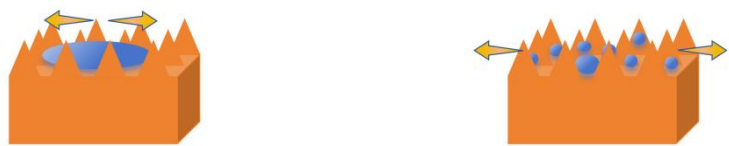

(b)

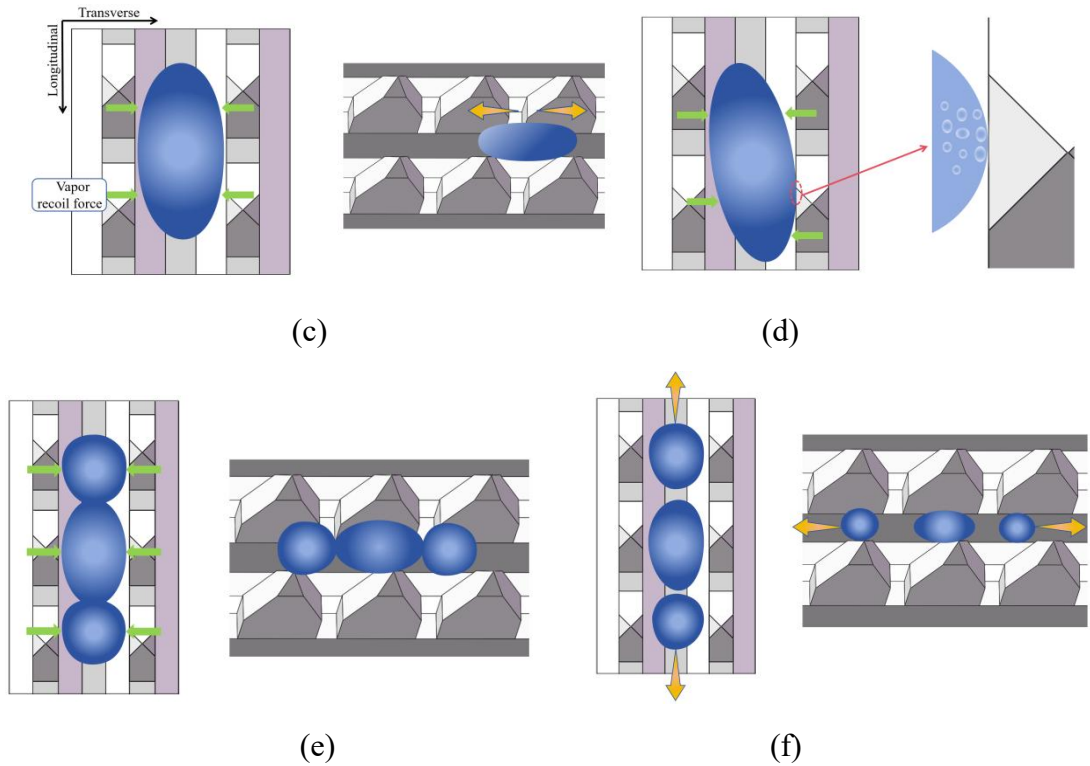

Figure 3　Dynamical states of droplets on spiked surfaces: (a) oscillation accompanied by volume reduction; (b) boiling and bursting; (c)-(f) the process of droplet migration. (c) Trapped between spikes; (d) contacting with the solid wall and boiling; (e) bursting into daughter droplets;(f) unidirectional migration.

Due to the geometric asymmetry of the spikes in transverse (width-wise) and longitudinal (length-wise) directions, the vapor recoil results in strong lateral compression in the transverse direction (Figure 3(e)), causing the droplet to burst into multiple daughter droplets that subsequently slide along the longitudinal direction (Figure 3(f)). However, in the transverse direction, the narrow inter-spike gaps generate significant resistance to droplet motion, effectively suppressing lateral displacement. This geometric asymmetry of the spiked surface is thus exploited to achieve controlled longitudinal droplet migration.

During droplet fragmentation, daughter droplets simultaneously move both leftward and rightward. The rightward-migrating droplets rebound upon colliding with the right baffle of the spiked zone, ultimately causing all droplets to eject outward through the left boundary of the spiked zone. This achieves unidirectional droplet migration (Figure 4(a) and video 6).

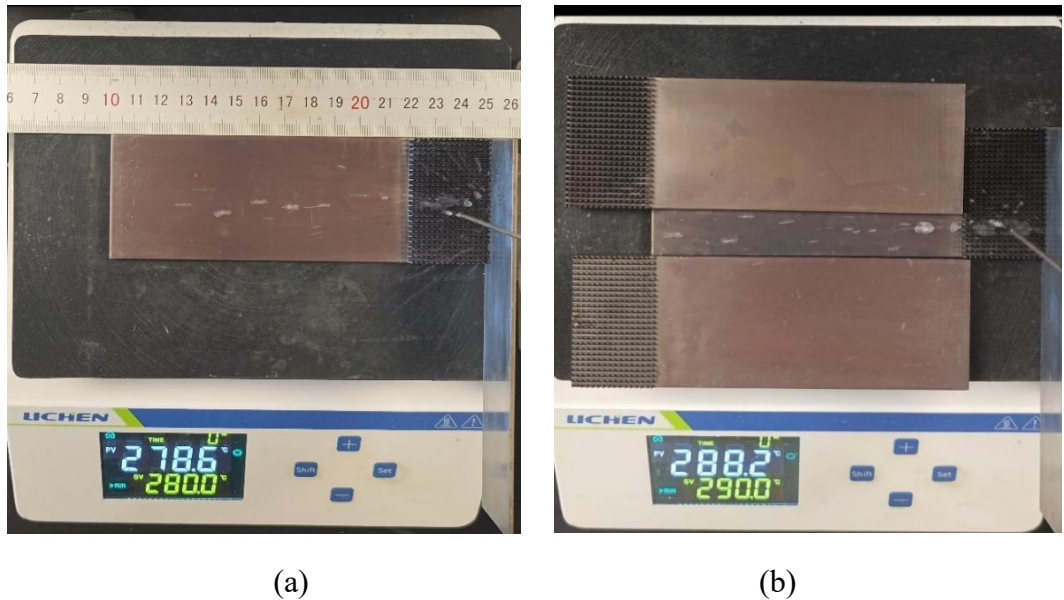

(a)            (b)

Figure 4    Unidirectional droplet migration: (a) on the plate and (b) within a narrower channel. Here, $h$=5mm.

When baffles are added along the width direction of the droplet migration region, the droplet migration can be confined within a narrower channel, as illustrated in Figure 4(b) and video 7. This suggests that we can design the channel width to control the migration zone width of the droplets.

3.2 Critical Migration Volume

We define the critical volume of a droplet at which migration occurs as the critical migration volume $V_m$. Figure 5 shows the variation of $V_m$ with the heating stage temperature $T$ and spike height $h$. The minimum droplet volume that the micropipette can produce is 3μL. It can be observed that $V_m$ increases with $T$ but decreases with $h$. This is because a higher solid temperature intensifies vaporization at the droplet's surface, leading to a thicker vapor layer between the droplet and the substrate. Consequently, the possibility of droplet contact with the solid wall is reduced, suppressing boiling and making it harder for the droplet to burst into smaller droplets or migrate. Increasing the volume of the droplet can enhance the probability of contact between the droplet and the solid wall. Conversely, increasing the spike height expands the area of droplet sidewall vaporization as well as the probability of droplet contact with the solid wall, enhancing boiling and promoting droplet fragmentation and migration.

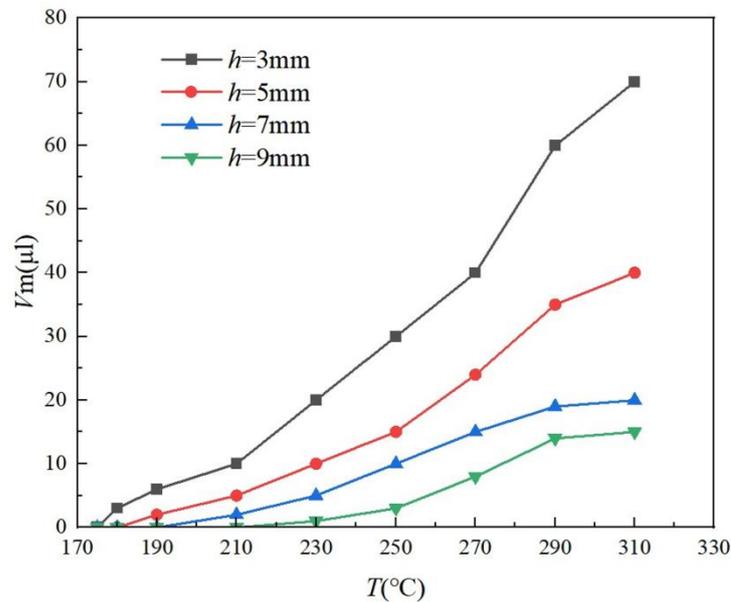

Figure 5　The variation of critical volume $V_m$ with the temperature of heating stage $T$ and spike height $h$. The droplets are injected using a 1μL precision micropipette.

For $h$=7mm and 9 mm, $V_m$ exhibits only a minor increase beyond 290°C as temperature rises. This phenomenon may be attributed to the fact that longer spikes exhibit higher thermal resistance. The temperature rise at the heating stage has a limited effect on the surface temperature of the spikes. As a result, the impact on droplet boiling behavior is reduced.

Furthermore, we identify that for the spiked plate with a specific spike height, there exists a critical temperature threshold $T_m$ for the heating stage. When the temperature falls below $T_m$, droplets of any size will only undergo boiling and cannot maintain a suspended state with oscillation (Figure 3(a) and video 4). For example, $T_m$=175°C, 180°C, 190°C and 210°C for $h$=3,5,7 and 9 mm, respectively, where the temperature is controlled to within ±1°C precision.

In addition, we can reasonably conclude that as the width of the microgroove ($L_1$ in Figure 1(b)) decreases, the surface area available for droplet vaporization at the sidewalls increases, thereby enhancing the likelihood of boiling and fragmentation. The above analysis demonstrates that the critical migration volume can be controlled by adjusting the spike height, microgroove width, and heating stage temperature.

3.3 Migration speed

Here, we measure the average speed of the droplet after it is ejected from the spiked region and moved across the flat area of the spiked plate, with a migration distance of 10 cm.

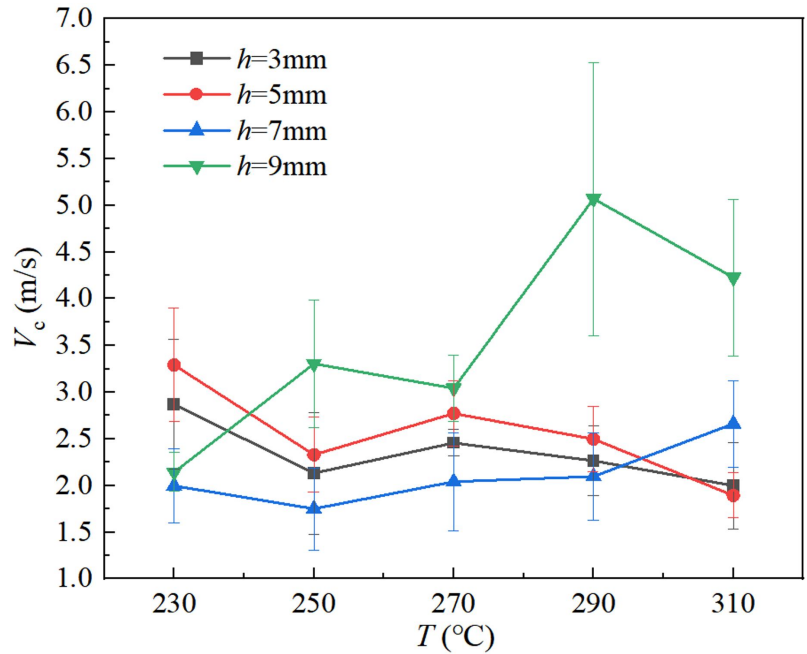

Figure 6　The variation of migration speed $V_c$ with $T$ and $h$. The volumes of droplets are 130μL.

Figure 6 shows the variation of droplet migration speed $V_c$ with the temperature of heating stage $T$ and spike height $h$, while the volume of all dispensed droplets remains constant at 130 μL. This ensures that the droplet volume exceeds the critical value $V_m$ across the entire temperature range (230°C-310°C) of the heating stage. The bars of uncertainty estimation represent the standard uncertainty.

It can be observed that $V_c$ exhibits non-monotonic dependence on both $T$ and $h$. Overall, the migration velocities under various conditions all remain within the same order of magnitude $O(3m/s)$, larger than the state-of-the-art droplet migration speeds in Table 1. The maximum migration velocity occurs when $h$=9mm and $T$ = 290°C . The average speed reaches $V_c$ =5.07 m/s, with one instance achieving a peak speed of 8.36 m/s (Figure 7 and video 8), significantly surpassing all previously reported maximum velocities of 1.10cm/s.

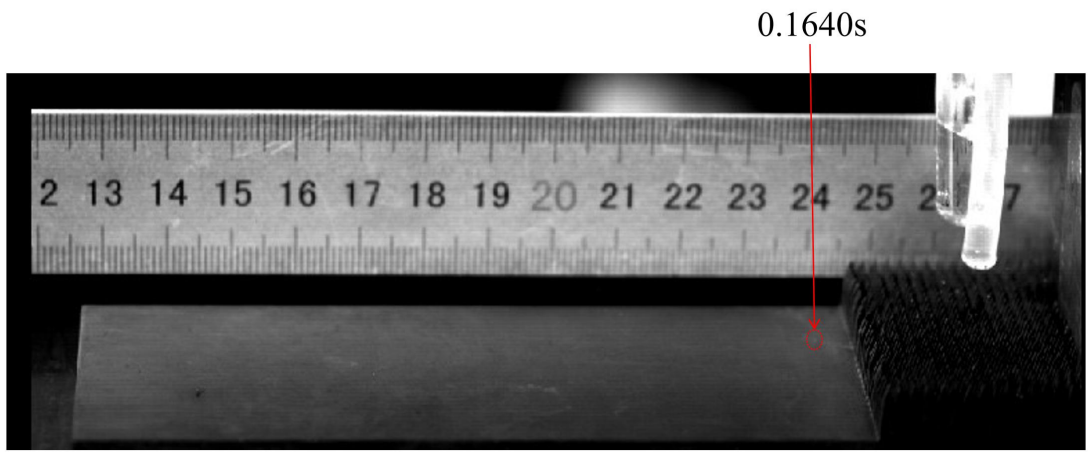

(a)

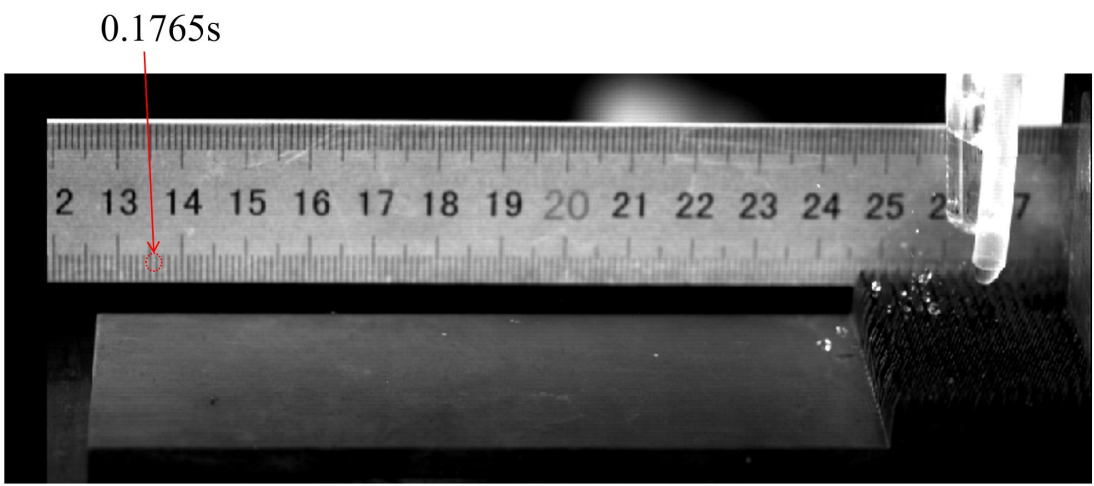

(b)

Figure 7　Droplet migration process with the fastest average speed at $h$=9mm and $T$ = 290°C: (a)　$t$=1.2065s and (b) $t$=1.2190s.

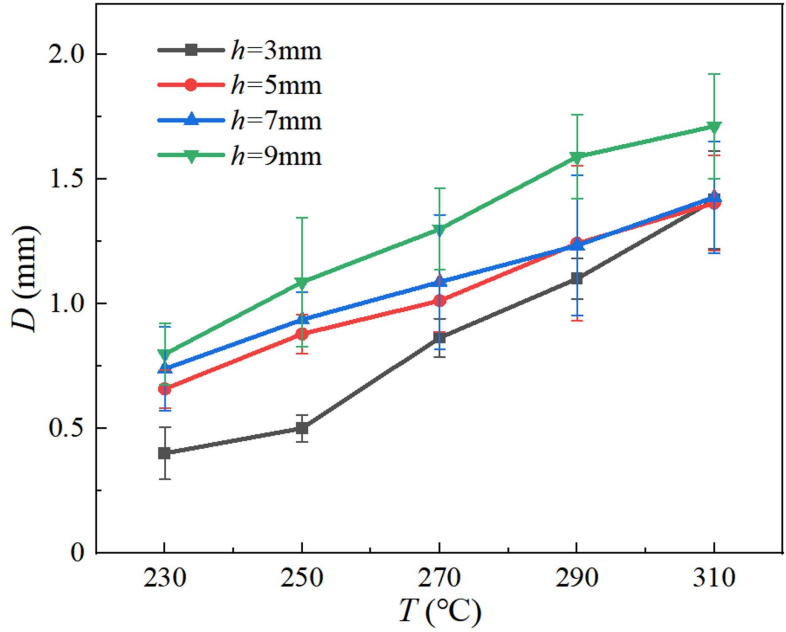

Figure 8　The variation of the migrating droplet diameter $D$ with $T$ and $h$.

Figure 8 shows the variation of the migrating droplet diameter $D$ with $T$ and $h$. The volumes of droplets are 130μL. It can be observed that $D$ generally increases with both $T$ and $h$. The reason can be explained as follows. When $T$ increases, boiling becomes more intense when the droplet contacts the solid wall, thereby generating larger daughter droplets through extrusion. Similarly, an increase in spike height enlarges the contact area between the droplet and the solid wall, resulting in more intense boiling and leading to an increase in the diameter of the daughter droplets.

　　The experiment shows that there is a certain correlation between the migration speed of liquid droplets and their diameter. Faster speeds are more commonly observed in smaller droplets. This is because during the bursting process, smaller droplets acquire a greater speed. We compare the distributions of $V_c$ and $D$ for daughter droplets at $T$ = 230°C, 310°C, as shown in Figure 9.

For $h$=3mm, the results demonstrate that when $T$ = 230°C, $V_c \in [2,3]$ m/s is the dominant range while $D$ mainly concentrates on 0-0.4 mm. However, when $T$ = 310°C, both $V_c$ and $D$ distributions show significantly greater dispersion.

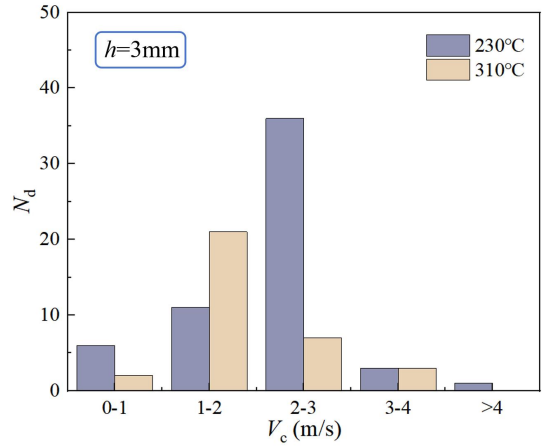

(a)

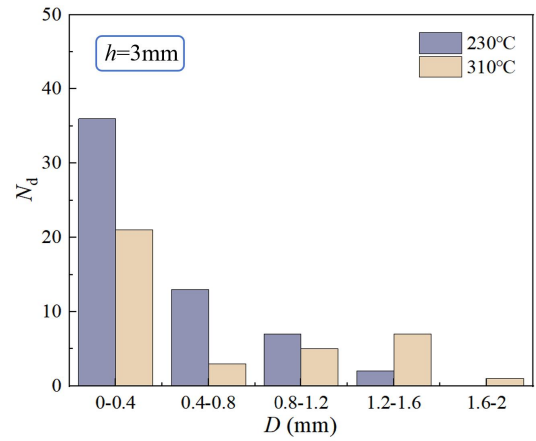

(b)

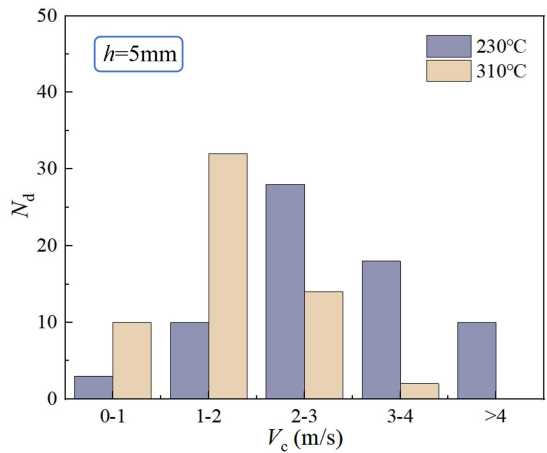

(c)

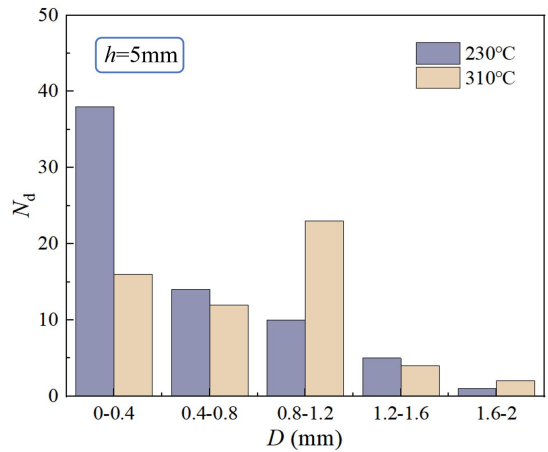

(d)

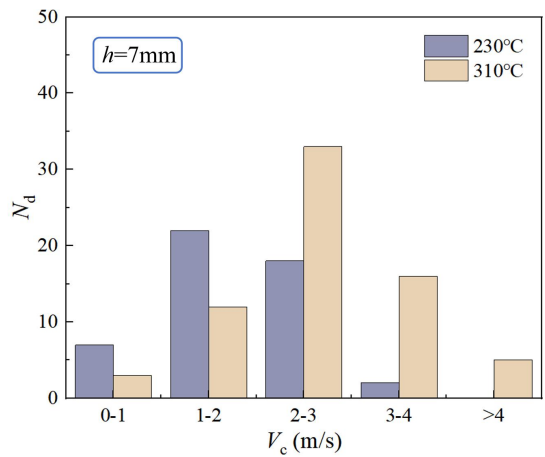

(e)

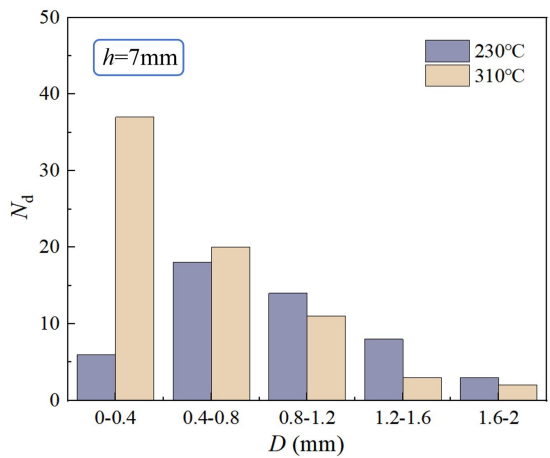

(f)

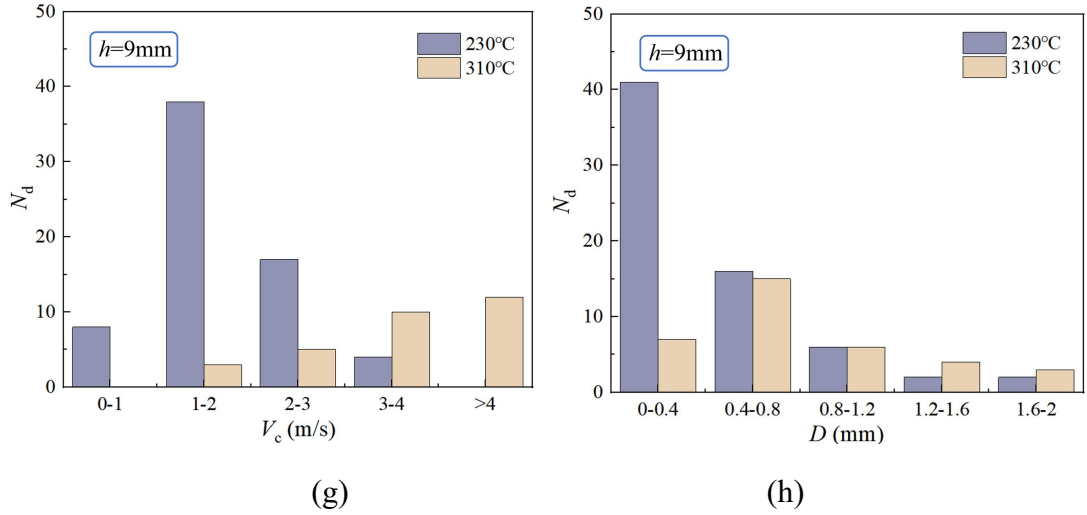

Figure 9  The distributions of $V_c$ and $D$ for daughter droplets at $T = 230°C, 310°C$: (a),(b) $h = 3$ mm; (c),(d) $h = 5$ mm; (e),(f) $h = 7$ mm; (g),(h) $h = 9$ mm.

For all other heights, droplets with moderate speeds generally dominate in proportion. The sole exception occurs at $h=9$mm, $T=310°C$, where the prevalence of slower droplets significantly decreases. From the perspective of size distribution, two primary patterns emerge. In the first type, smaller droplets account for the largest proportion, with droplet prevalence decreasing as size increases. In the second type, medium-sized droplets dominate at a higher proportion.

3.4  The case at $h=3$mm

We find that the case at $h=3$mm is the most complex, where droplets can exhibit three distinct states. In the small-volume regime, the droplet cannot encapsulate the spike and instead settles into the microgroove, where it remains suspended and undergoes evaporation (Figure 2(b)). In the intermediate-volume regime, the droplet can undergo boiling-driven bursting within the microgroove (Figure 2(c)). In the large-volume regime, in addition to boiling-induced bursting, the droplet may also encapsulate the spike (Figure 2(a)). Over time, the droplet encapsulating the spike gradually decreases in volume, subsequently settles into the microgroove, and undergoes boiling-induced bursting.

## 4. Conclusion

In summary, we achieve rapid directional transport of Leidenfrost droplets using a spiked substrate with a maximum average speed reaching 8.36 m/s over a 10 cm distance, significantly surpassing all previously reported maximum values. The fundamental mechanism exploits the instability of gas film between the solid wall and the droplet.

Upon heating the substrate above the Leidenfrost temperature $T_\mathrm{L}$, droplets confined between spikes exhibit levitation. The sides and bottom of the droplet undergo vaporization, generating a gas film. When this film becomes unstable and ruptures, the droplets come into direct contact with spike surfaces, triggering violent boiling that strongly compresses and fragments them into daughter droplets, which then undergo rapid directional transport. The geometric asymmetry of the spiked shape causes the droplet to move only along the longitudinal direction. The baffle on one side of the spiked region causes all droplets to ultimately move in a single direction.

There exist three distinct Leidenfrost states. In the first scenario, the droplet envelops the spikes, occurring only when the spikes are relatively short. In the second scenario, the droplet levitates and oscillates between the spikes. When the droplet volume exceeds a critical threshold, the third scenario occurs: the droplet undergoes boiling, splitting into multiple daughter droplets that rapidly migrate in a directional manner. The last two scenarios can be observed across four different spike-plate lengths. The speed of droplet migration exhibits a non-monotonic relationship with both the temperature of heating stage and the spike height, whereas the droplet diameter increases monotonically with both parameters.

The spiked copper substrates were used in our previous condensation experiment [23]. In this work, the vertical alignment of the spikes subjects the droplet to asymmetric lateral pressure, facilitating contact with the solid wall where boiling and bursting occur. The geometric asymmetry of the spikes (vertical vs. horizontal) restricts droplet migration to the vertical direction only. This suggests that the spiked surface also holds significant value in droplet migration.

Appendix

1.Preparation of Spiked Copper substrate

The preparation of spiked copper substrates is completed by Zhuangshi Liquanya Machinery Hardware Factory in Ningbo Zhenhai District using Beijing Andejianqi AR35MA precision CNC electric spark wire cutting machine (machine tool processing accuracy ±0.001). The parameters of the substrate in Figure 1 are displayed in Table 2.

**Table 2  The parameters the substrate in Figure 1**

| $L_1$ | $L_2$ | $L_3$ | $L_4$ | $L_5$ | $L_6$ | $L_7$ |
|---|---|---|---|---|---|---|
| 1.5 | 0.5 | 0.54 | 0.5 | 1 | 1.61 | 3 |

All values are given in mm.

After the preparation, we place the copper plate in an ethanol solution and clean it in an ultrasonic cleaner (LiChen brand) for 5 minutes. After drying, we immerse the copper plate in a dilute sulfuric acid solution (0.5 mol/L) to remove surface oxides, then rinse with deionized water for 5 minutes. For smooth surfaces(transport zone in Figure 1), sandpapers with grit sizes of 500, 1000, 1500, and 2000 are sequentially used for polishing. Subsequently, the polished substrate is placed in an ultrasonic cleaner containing ethanol solution to remove surface grease and other impurities. After 10 minutes of cleaning, the substrate is removed. Finally, the treated substrate is rinsed with high-purity water and dried in an electric blast drying oven at 100°C.

In the experiment, the copper substrate undergoes surface oxidation under high temperatures. In Figure 10, we compare SEM (scanning electron microscopy) images of the copper surface before and after oxidation.

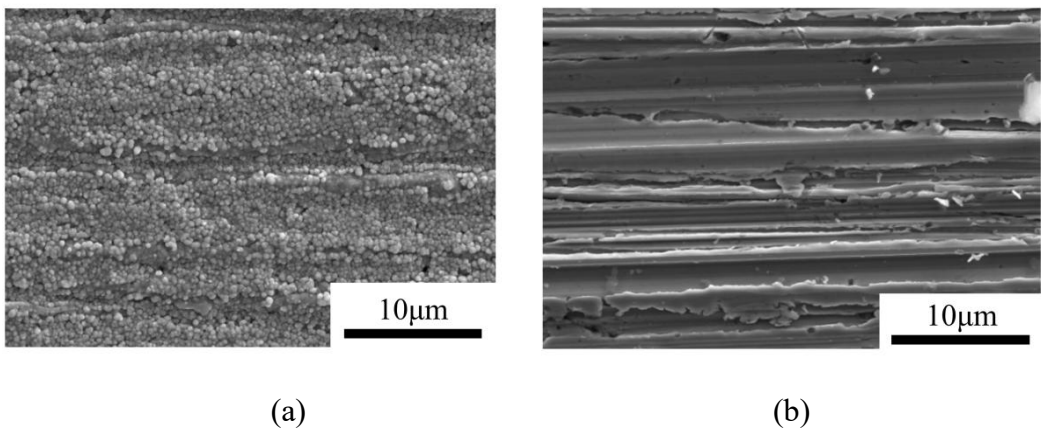

(a)　　　　　　　　　　　　　　(b)

Figure 10　SEM images of the copper surface (a) before and (b) after oxidation.

2.Instrument and Materials

Table 3 lists the models and origins of the instruments, while Table 4 provides details on the materials, specifications and manufacturers used in the experiment.

Table 3　　Instruments

| Instrument name | Model | Manufacturer |
|---|---|---|
| Electric blast drying oven | LC-101-00BS | Lichchen |
| Ultra pure water machine | UPT-11-10T | Youpu |
| Ultrasonic cleaner | PS-031ST | ULTRASONIC CLEANER |
| Scanning electron microscopy | SU5000 | Hitachi |
| High-speed camera | FASTCAM SA1.1 | Photron |
| Heating Stage | LC-DB-XDA | Lichchen |
| Pipette | 0.1-10 μL/10-100 μL | Sanliang |
| Peristaltic pump | TS006GMYZ-05 | Tushui |

Table 4　　Materials

| Material, | Specification | Manufacturer |
|---|---|---|
| Absolute ethyl alcohol | AR，500 mL/bottle | Xilong |
| Polyacrylamide particles | AR，1 kg/bottle | Xilong |
| HF-52 | AR，5 kg/bottle | Ziyun |
| T2 Purple copper plate | 150 mm×30 mm×10mm | Dongguan Yuxin Metal Materials Co., LTD |

**Data availability**

The authors declare that the data supporting the findings of this study are available within the paper and its supplementary information files. Source data are available upon reasonable request.

**Declaration of Interests**

The authors declare no competing interests.

**Author contributions**

Kai-xin Hu made substantial contributions to the conception of the work, wrote the paper for important intellectual content and approved the final version to be published. He is accountable for all aspects of the work in ensuring that questions related to the accuracy or integrity of any part of the work are appropriately investigated and resolved. Yin-Jiang Chen and Dong-Xu Duang conducted experiments, prepared the materials and drew the figures. Dan Wu provided experimental equipment and funding. Qi-shen Chen provided editing and writing assistance.

**Acknowledgments**

This work has been supported by the National Natural Science Foundation of China (No.12372247), Zhejiang Provincial Natural Science Foundation (No.LZ25A020009), Ningbo Municipality Key Research and Development Program (No. 2022Z213) and the China Manned Space Engineering Application Program－China Space Station Experiment Project (No. TGMTYY1401S).

**References**

[1] Lv, P. , Yong‐Lai Zhang, Dong‐Dong Han, & Hong‐Bo Sun. (2021). Directional droplet transport on functional surfaces with superwettabilities. Advanced Materials Interfaces, 2100043.

[2] Liu, C., Sun, Y., Huanng, J., Guo, Z., & Liu, W. . (2021). External-field-induced directional droplet transport: a review. Advances in Colloid and Interface Science, 295, 102502.

[3] Lin S, Li B, Xu Y, et al. Effective Strategies for Droplet Transport on Solid Surfaces[J].Advanced Materials Interfaces, 2021, 2001441.

[4] Wu L, Guo Z, Liu W. Surface behaviors of droplet manipulation in microfluidics devices[J]. Advances in Colloid and Interface Science, 2022: 102770.

[5] Shastry A, Case M J, Böhringer K F. Directing droplets using microstructured surfaces[J]. Langmuir, 2006, 22(14): 6161-6167.

[6] Dai Q, Huang W, Wang X. Surface roughness and orientation effects on the thermo-capillary migration of adroplet of paraffin oil[J]. Experimental thermal and fluid science, 2014, 57: 200-206.

[7] Linke H, Alemán B J, Melling L D, et al. Self-propelled Leidenfrost droplets[J]. Physical review letters, 2006, 96(15): 154502.

[8] Kruse C, Somanas I, Anderson T, et al. Self-propelled droplets on heated surfaces with angled self-assembled micro/nanostructures[J]. Microfluidics and nanofluidics, 2015, 18: 1417-1424.

[9] Li J, Hou Y, Liu Y, et al. Directional transport of high-temperature Janus droplets mediated by structural topography[J]. Nature Physics, 2016, 12(6): 606-612.

[10] Sobac B, Rednikov A, Dorbolo S, et al. Self-propelled Leidenfrost drops on a

thermal gradient: A theoretical study[J]. Physics of Fluids, 2017, 29 (8): 082101.

[11] Lyu S, Tan H, Wakata Y, et al. On explosive boiling of a multicomponent Leidenfrost drop[J]. Proceedings of the National Academy of Sciences, 2021, 118(2): e2016107118.

[12] Graeber G, Regulagadda K, Hodel P, et al. Leidenfrost droplet trampolining[J]. Nature communications, 2021, 12(1): 1727.

[13] Chen H, Zhang P, Zhang L, et al. Continuous directional water transport on the peristome surface of Nepenthes alata[J]. Nature, 2016, 532(7597): 85-89.

[14] Style R W, Che Y, Park S J, et al. Patterning droplets with durotaxis[J]. Proceedings of the National Academy of Sciences, 2013, 110(31): 12541-12544.

[15] Daniel S, Chaudhury M K, Chen J C. Fast drop movements resulting from the phase change on a gradient surface[J]. Science, 2001, 291(5504): 633-636.

[16] Sun Q, Wang D, Li Y, et al. Surface charge printing for programmed droplet transport[J]. Nature materials, 2019, 18(9): 936-941.

[17] Quéré D. Leidenfrost dynamics[J]. Annual Review of Fluid Mechanics, 2013, 45(1): 197-215.

[18] Mura, E., Massoli, P., Josset, C., Loubar, K., & Bellettre, J. (2012). Study of the micro-explosion temperature of water in oil emulsion droplets during the Leidenfrost effect. Experimental Thermal and Fluid Science, 43, 63-70.

[19] Vakarelski I U, Berry J D, Chan D Y C, et al. Leidenfrost vapor layers reduce

drag without the crisis in high viscosity liquids[J]. Physical Review Letters, 2016, 117(11): 114503.

[20] Wells G G, Ledesma-Aguilar R, McHale G, et al. A sublimation heat engine. Nature Communications, 2015, 6(1): 1-7.

[21] Wang K, Zhang H, Wang Y, et al. Power generation from an elastic Leidenfrost hydrogel piston enabled heat engine. International Journal of Heat and Mass Transfer, 2021, 179: 121661.

[22] Dupeux G, Baier T, Bacot V, et al. Self-propelling uneven Leidenfrost solids[J]. Physics of Fluids, 2013, 25 (5): 051704.

[23] Hu KX, Chen YJ, Wu D, Tang BW, Tao YQ, Chen QS. Efficient condensation on spiked surfaces with superhydrophobic and superhydrophilic coatings. International Journal of Heat and Mass Transfer. 2025 , 246:127055.